\documentclass{article}

\usepackage{lineno,hyperref}

\usepackage{pifont}
\usepackage{tikz}
\usetikzlibrary{positioning}

\begin{document}

\begin{center}
   \large
   \textbf{On the Simulation of Hypervisor Instructions for Accurate Timing Simulation of Virtualized Systems}
   
   \normalsize
   \vspace{0.7cm}
   Swapneel C. Mhatre\\
   Department of Computer Science and Engineering\\
   National Institute of Technology Calicut\\
   Kerala, India\\
   mhatreswapneel\_p170067cs@nitc.ac.in\\
   \vspace{0.7cm}
   Priya Chandran\\
   Department of Computer Science and Engineering\\
   National Institute of Technology Calicut\\
   Kerala, India\\
   priya@nitc.ac.in\\
   \vspace{0.7cm}
\end{center}

\begin{abstract}
Architectural simulators help in better understanding the behaviour of existing architectures and the design of new architectures. Virtualization has regained importance and this has put a pressing demand for the simulation of virtualized systems. Existing full-system simulators for virtualized systems simulate the application program instructions and the operating system instructions but abstract the hypercalls or traps to the hypervisor. This leads to inaccuracy in the simulation. This paper proposes an approach to simulate hypervisor instructions in addition to operating system instructions for accurate timing simulation of virtualized systems. The proposed approach is demonstrated by simulating RISC-V binary instructions. The simulator is an execution-driven, functional-first, hardware-based simulator coded in Verilog. The paper concludes that the proposed approach leads to accurate timing simulation of virtualized systems.
\end{abstract}


\section{Introduction}
Architectural simulators help in better understanding the behaviour of existing architectures and the design of new architectures. While a prototype is an exact replica of the system being modeled, a simulator imitates the modeled system to the desired level of details in the desired time \cite{Hari2014}. Prototypes, though far more accurate than simulators, are far slower than the simulators. Hence, simulators play a vital role in modeling large computer systems in desired time. The system modeled by the simulator is called the \textit{target}, while the system that executes or implements the simulator is called the \textit{host} \cite{Hari2014}. The host may be a computer or a field-programmable gate array (FPGA).

Virtualization has regained importance and this has put a pressing demand for the simulation of virtualized systems, especially, those with security-related enhancements \cite{Mhatre2018}. However, it is a challenging task to simulate virtualized systems and hence, there is a dearth of simulators for such systems. The challenges are mainly due to the increased overhead due to virtualization in a virtualized system. In a \textit{non-virtualized system}, the central processing unit (CPU) and memory are virtualized, but input/output (I/O) is not virtualized. However, in a \textit{virtualized system}, CPU, memory as well as I/O are virtualized. Furthermore, the overhead due to CPU and memory virtualization in a virtualized system is more than that in a non-virtualized system. This makes the simulators for non-virtualized systems unsuitable for simulating virtualized systems.

In order to accurately simulate the overhead due to virtualization, it is necessary for a \textit{full-system simulator} to simulate the hypervisor (HV) instructions in addition to the operating system (OS) instructions. However, this increases the simulation time \cite{Nikounia2015}. Hence, existing simulators like \textit{Virtual-GEMS} \cite{Antonio2009} and \textit{gem5v} \cite{Nikounia2015} abstract the hypercalls or traps to the hypervisor code from the OS code. This leads to inaccuracy in the timing simulation of virtualized systems.

Driven by the urgent need for the simulation of virtualized systems and the lack of accurate simulators for such systems, this paper proposes a hardware-based approach for the simulation of hypervisor instructions. Software is inherently sequential, while hardware is inherently parallel. Hence, hardware-based simulator is faster than the software-based simulator and can simulate the increased overhead due to virtualization in reasonable time. The simulator code is written using Verilog so as to develop a hardware-based simulator. 

Since RISC-V \cite{Krste2014, Andrew2016, Tony2016, Andrew2020-I} is an open standard instruction set architecture (ISA), the approach is demonstrated by simulating RISC-V binary instructions. The simulator is a \textit{decoupled simulator} \cite{Mauer2002}, particularly, \textit{functional-first simulator} \cite{Hari2014} and is an \textit{execution-driven simulator} \cite{Akram2016}.

The contributions of the paper are:
\begin{enumerate}
\item An approach for the simulation of hypervisor instructions in virtualized systems
\item Demonstration of the approach by simulating RISC-V program binary
\end{enumerate}

Section 2 gives the capabilities of existing simulators for virtualized systems. Although few simulators for virtualized systems exist, they have their own limitations, the major limitation being the lack of timing accuracy in simulating the increased overhead due to virtualization. Section 3 gives the proposed approach for the simulation of hypervisor instructions in virtualized systems. Section 4 demonstrates the proposed approach by simulating RISC-V binary instructions since RISC-V is an open standard ISA. Section 5 gives the simulator configuration. Section 6 gives the results obtained while running searching and sorting programs. Section 7 gives the way in which the simulator is validated. Section 8 concludes that the proposed approach leads to accurate timing simulation of virtualized systems. Section 9 gives future directions for research.

\section{Related work}
\textit{Multikernel simulation} \cite{Chandran2017, Aditya2016} is an interesting approach to simulate rollback-sensitive memory architecture. It simulates the hardware modifications by modifying the kernels at different privilege levels. It modifies the hypervisor and the guest OS and if required, even the guest application program. However, multikernel simulation performs only functional simulation and not timing simulation.

An interesting approach to simulate virtualization-enabled processors effectively is to extract the required information from the hypervisor like detection of the hypercalls and memory accesses \cite{Mhatre2020}. However, this requires modifications to be done to the hypervisor source code.

\textit{PVMsim} \cite{Wu2010}, based on \textit{Multi2Sim} \cite{Ubal2007, Ubal2012, Chris}, is an \textit{application-level} \textit{functional} and \textit{timing simulator}. PVMsim simulates virtualized systems. However, being an application-level simulator, it does not simulate OS instructions, but abstracts system calls. Furthermore, it does not even simulate hypervisor instructions. Whenever the application program encounters a system call, PVMsim emulates it as a single ISA instruction and just performs the hypervisor operations in response to the system call without actually simulating the OS and hypervisor instructions that get executed on the system call. This leads to large inaccuracy in the timing simulation of virtualized systems.

\textit{Virtual-GEMS} \cite{Antonio2009}, based on \textit{Simics} \cite{Magnusson2002} and \textit{GEMS} \cite{Martin2005}, is a \textit{full-system} \textit{functional} and \textit{timing simulator}. \textit{gem5v} \cite{Nikounia2015}, a modified version of \textit{gem5} \cite{Binkert2011}, is also a \textit{full-system} \textit{functional} and \textit{timing simulator}. Both Virtual-GEMS and gem5v simulate virtualized systems. Being full-system simulators, they simulate OS instructions and abstract I/O. However, they do not simulate hypervisor instructions. They abstract the hypercalls, or traps to the hypervisor code from the OS code. Every hypercall or trap to the hypervisor code from the OS code is emulated as a single ISA instruction. They implement the functionality of the hypervisor, but do not give the timing overhead introduced by the hypervisor. As Virtual-GEMS and gem5v simulate OS instructions, their timing accuracy is higher than that of PVMsim, nevertheless, since the hypervisor instructions are not simulated, their accuracy is far from being acceptable in many practical situations.

\textit{Mambo} \cite{Bohrer2004} is a \textit{full-system} \textit{functional} and \textit{timing simulator}. It can model hypervisor support. However, it does not simulate hypervisor instructions.

\textit{PTLsim} \cite{Yourst2007} is a \textit{timing simulator}. The userspace version of PTLsim is integrated with Xen hypervisor \cite{Barham2003, Clark2004} to form the full system version of PTLsim, called \textit{PTLsim/X}. Although PTLsim (full system version) uses Xen virtualization facilities, it does not simulate hypervisor instructions. 

\textit{QEMU (Quick EMUlator)} \cite{Bellard2005} is a \textit{hypervisor} as well as a \textit{functional simulator} that supports both \textit{Linux user mode emulation} and \textit{full system emulation}. But it is not a timing simulator. It does not simulate hypervisor instructions.

\textit{MARSS (Micro ARchitectural and System Simulator)} \cite{Patel2011} is a \textit{full-system} \textit{functional} and \textit{timing simulator} that modifies \textit{PTLsim} \cite{Yourst2007} (full system version) and combines it with \textit{QEMU} \cite{Bellard2005}. Although PTLsim is a timing simulator, it does not simulate hypervisor instructions and QEMU is just a functional simulator. Since MARSS uses QEMU to decode and emulate complex opcodes including system calls and returns and PTLsim to simulate the interrupt handlers, MARSS does not simulate hypervisor instructions.

\textit{VMcSim} \cite{Tchana2015}, based on \textit{McSimA+} \cite{Ahn2013}, is a \textit{functional} and \textit{timing simulator}. VMcSim simulates virtualized systems. Like PVMsim, it does not simulate OS instructions. However, it simulates simulated hypervisor instructions, although it does not simulate real hypervisor instructions. Furthermore, VMcSim simulates only those simulated hypervisor instructions that perform scheduling.

\textit{PVMsim} \cite{Wu2010}, \textit{Virtual-GEMS} \cite{Antonio2009}, \textit{gem5v} \cite{Nikounia2015}, \textit{Mambo} \cite{Bohrer2004}, \textit{PTLsim} \cite{Yourst2007}, \textit{QEMU} \cite{Bellard2005} and \textit{MARSS} \cite{Patel2011} do not model the increased overhead in CPU virtualization and memory virtualization in virtualized systems. \textit{VMcSim} \cite{Tchana2015} models the increased overhead in CPU virtualization, but does not model the increased overhead in memory virtualization in virtualized systems.

A comparison of the capabilities of various simulators for virtualized systems is given in Table \ref{Capabilities_of_Various_Simulators_for_Virtualized_Systems}.

\begin{table}
\tiny
   \caption{Capabilities of Various Simulators for Virtualized Systems}
   \label{Capabilities_of_Various_Simulators_for_Virtualized_Systems}
   \begin{tabular}{cccccccc}
      \hline 
      Simulator & Timing        & Timing        & Timing        & Ability to     & Ability to & Ability to\\
                & Simulation of & Simulation of & Simulation of & Model          & Model & Model\\
                & Application   & OS            & Hypervisor    & the Increased  & the Increased & the Increased\\
                & Program       & Instructions  & Instructions  & Overhead in    & Overhead in & Overhead in\\
                & Instructions  &               &               & CPU            & Memory & I/O\\
                &               &               &               & Virtualization & Virtualization & Virtualization\\
                &               &               &               & in Virtualized & in Virtualized & in Virtualized\\
                &               &               &               & Systems        & Systems & Systems\\
      \hline 
      PVMsim                  & \ding{51} & \ding{53} & \ding{53} & \ding{53} & \ding{53} & \ding{53}\\\\
      Virtual-GEMS            & \ding{51} & \ding{51} & \ding{53} & \ding{53} & \ding{53} & \ding{53}\\\\
      gem5v                   & \ding{51} & \ding{51} & \ding{53} & \ding{53} & \ding{53} & \ding{53}\\\\
      Mambo                   & \ding{51} & \ding{51} & \ding{53} & \ding{53} & \ding{53} & \ding{53}\\\\
      PTLsim                  & \ding{51} & \ding{51} & \ding{53} & \ding{53} & \ding{53} & \ding{53}\\
      (Full system version)   &           &           &           &           &           &\\\\
      QEMU                    & \ding{53} & \ding{53} & \ding{53} & \ding{53} & \ding{53} & \ding{53}\\
      (Full system emulation) &           &           &           &           &           &\\\\
      MARSS                   & \ding{51} & \ding{51} & \ding{53} & \ding{53} & \ding{53} & \ding{53}\\\\
      VMcSim                  & \ding{51} & \ding{53} & *         & \ding{51} & \ding{53} & \ding{53}\\\\
      \hline 
      \multicolumn{7}{l}{* Simulates simulated hypervisor instructions, but not real hypervisor instructions,}\\
      \multicolumn{6}{l}{furthermore, simulates only those simulated hypervisor instructions that perform scheduling}\\
   \end{tabular}
\end{table}

The main limitation of the existing simulators is the lack of the simulation of hypervisor instructions in virtualized systems. Simulating an OS itself takes a long time. Simulating a hypervisor along with the OS would further exacerbate the time required for simulation. If the simulation time becomes excessively high, it would drive the simulator practically useless. Hence, to reduce the simulation time, a hardware-based approach is proposed to simulate the hypervisor instructions in virtualized systems. 

\section{Proposed approach}
Software-based simulators run at the clock frequency of the computers on which they are installed, while hardware-based simulators run at the clock frequency of the FPGA on which they are downloaded. Although the clock frequency of the computers is higher than the clock frequency of the FPGA, the software is inherently sequential, while the hardware is inherently parallel. Hardware-based simulators can be divided into independent modules that can run in parallel, while the independent modules of software-based simulators would run in pseudo-parallel manner unless they are run on multicore processors. Even if they are run on multicore processors, the number of modules that can run in parallel is limited by the number of cores in the processors. Hence, hardware-based simulators are faster than the corresponding software-based simulators and can simulate the increased overhead due to virtualization in reasonable time. The proposed approach uses Verilog, which is a Hardware Description Language (HDL), to write the simulator code so as to develop a hardware-based simulator. 

The simulator maintains the \textit{state of the program} and the \textit{state of the hardware}. The \textit{state of the program} consists of \textit{architected register file} and \textit{virtual memory image}. The \textit{architected register file} includes the unprivileged and privileged registers. The unprivileged registers include the general-purpose registers and program counter (PC) or instruction pointer (IP) that are accessible from the non-privileged and privileged modes. The privileged registers include the control and status registers (csrs) that are accessible only from the privileged mode. The \textit{virtual memory image} consists of different portions of the addressable memory corresponding to the privilege levels or modes supported by the processor. Each portion is further divided into areas for code, data and stack. The \textit{state of the hardware} consists of the privilege level or mode of the processor.

The application program binary code, the OS binary code and the hypervisor binary code is stored in the code areas of the corresponding portions of the memory. The OS code includes the trap handlers that handle the system calls. The trap handlers include the code for context switching and the system call service routines. Likewise, the hypervisor code includes the trap handlers that handle the hypercalls. The page tables are also stored in the memory to simulate memory virtualization.

When the simulator is reset, the state of the program and the state of the hardware may be initialized to the state when the processor is reset. However, in order to bypass the instructions of the OS that are executed in the beginning when the computer is booted, the state of the program and the state of the hardware is initialized to the state when the processor comes to the program entry point. Whenever the simulator encounters a system call in the non-privileged mode (user space), it updates the PC or IP so as to point to the location in the privileged mode (kernel space) as dictated by the architecture of the processor. This simulates a jump from the user space (user land) to the kernel space (kernel land) and simulates the system call. Similarly, when a hypercall is encountered while simulating the OS code, the simulator updates the PC or IP so as to point to the location where the hypervisor code resides. This exact location is also dictated by the architecture of the processor. This simulates a jump from the OS code to the hypervisor code and simulates the hypercall. Return-from-trap is also simulated in a similar way by proper updation of the PC or IP.

The proposed approach is demonstrated by simulating RISC-V program binary.

\section{Demonstration of the proposed approach}
Since RISC-V \cite{Krste2014, Andrew2016, Tony2016, Andrew2020-I} is an open standard ISA, the proposed approach is demonstrated by simulating RISC-V binary instructions, particularly, RV32I, which is one of the four base integer ISAs in RISC-V. The other base ISAs, viz., RV64I, RV32E and RV128I and the optional extensions to the base ISA are not simulated. (`I' in RV32I, RV64I and RV128I stands for `integer', while `E' in RV32E, which is a subset variant of RV32I, stands for `embedded'.) RV32I is characterized by:
\begin{enumerate}
\item 32-bit integer registers
\item 32 integer registers other than the program counter in unprivileged mode
\item $ 2^{32} $ bytes = 4 GiB byte-addressable address space
\end{enumerate}

The simulator implements \textit{hybrid virtualization}, which is a technique of combining \textit{paravirtualization} (modifying the guest OS so that it uses hypercalls) with \textit{hardware-assisted virtualization} (introducing additional privilege levels or modes so that the hypervisor can be at a higher privilege level than the OS without evicting the OS from the level where it is intended to be). RISC-V implements hardware-assisted virtualization using the \textit{RISC-V Hypervisor Extension (H-Extension)} \cite{Andrew2020-II}. With the H-Extension, there are two modes for guest execution - Virtualized Supervisor mode (VS-mode) and Virtualized User mode (VU-mode). The VS-mode is intended for the guest operating system, while the VU-mode is intended for the guest user programs. Moreover, with the H-Extension, some new functionality is added to the supervisor mode (S-mode) so that it can access the hypervisor and the S-mode becomes Hypervisor-Extended Supervisor mode (HS-mode) or Hypervisor mode. The HS-mode is intended for the hypervisor in virtualized systems and for the OS in non-virtualized systems. Thus, the OS always runs virtualized.

The environment call in RISC-V is a call to the supporting execution environment. The supporting execution environment may be operating system, in which case the environment call is a system call, or hypervisor, in which case the environment call is a hypercall. RISC-V can make an environment call from any mode and by default, the trap goes to the machine mode (M-mode), the highest privilege mode in RISC-V. However, the trap may be delegated to the lower privilege modes using the trap delegation registers, which are the csrs available in RISC-V. The simulator may be configured to simulate virtualized or non-virtualized systems. If the simulator is configured to simulate virtualized systems, it assumes that the environment call from the VU-mode is delegated to the VS-mode, while the environment call from the VS-mode is delegated to the HS-mode. If the simulator is configured to simulate non-virtualized systems, it assumes that the environment call from the user mode (U-mode), which is essentially the VU-mode in virtualized systems, is delegated to the HS-mode. RISC-V uses direct traps for synchronous exceptions, i.e., all the synchronous exceptions trap to a specific address and the cause registers, which are the csrs in RISC-V, determine the cause of the exception. Accordingly, the simulator simulates ECALL (environemnt call) instructions using direct traps (but does not simulate any asynchronous interrupts in the present implementation).

The simulator code is written using Verilog so as to develop a hardware-based simulator.

\begin{figure}
\centering

\includegraphics[width=0.97\textwidth]{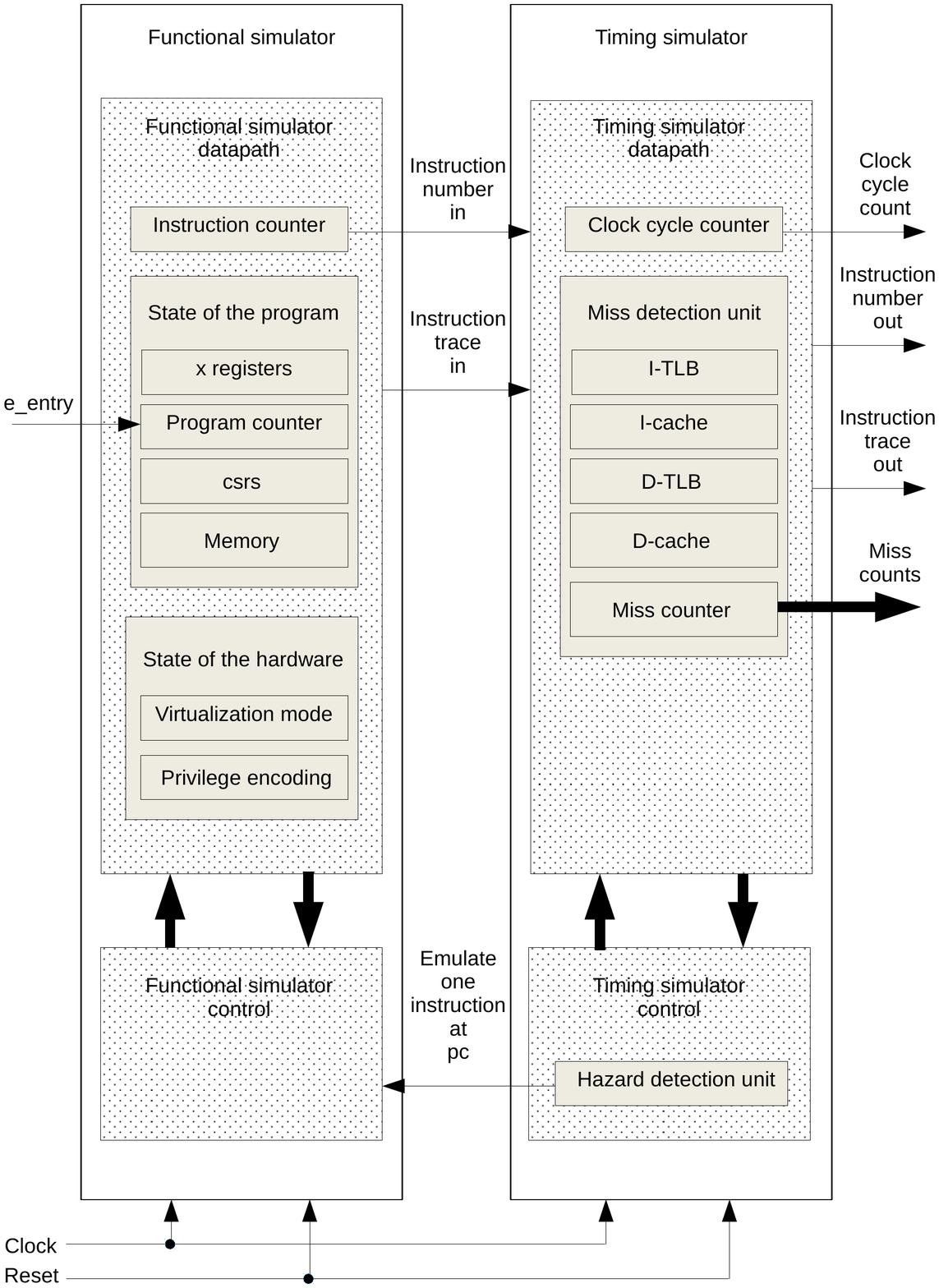}

\caption{Simulator.}
\label{Simulator}
\end{figure}

The simulator is a \textit{decoupled simulator} \cite{Mauer2002}, i.e., the \textit{functional simulator} (that implements the architecture) and the \textit{timing simulator} (that implements the microarchitecture) are decoupled in separate codes. Separation of functional simulation component from the timing simulation component eases simulator development and modification. If there are changes only in the ISA, only the functional simulation component needs to be modified. If there are changes only in the microarchitecture, only the timing simulation component needs to be modified. There are different types of decoupled simulators based on whether functional simulation is done first or timing simulation is done first or whether the functional simulator directs the timing simulator or the timing simulator directs the functional simulator. Thus, there are four types of decoupled simulators, viz., \textit{functional-first simulators} \cite{Hari2014}, \textit{timing-directed simulators} \cite{Hari2014}, \textit{timing-first simulators} \cite{Hari2014, Mauer2002} and \textit{functional-directed simulators} \cite{Ryckbosch2010}. The proposed simulator is a \textit{functional-first simulator}. In a \textit{functional-first simulator}, the functional simulator generates the instruction traces on-the-fly and updates the state of the program and a separate timing simulator accepts the instruction traces generated by the functional simulator and generates the performance statistics. The simulator is an \textit{execution-driven simulator} \cite{Akram2016}, i.e., it directly uses the program instructions in binary to perform the simulation, generating the instruction traces on-the-fly. This eliminates the need for huge amount of memory that is required by \textit{trace-driven simulators} \cite{Akram2016} to store the instruction traces. The block diagram of the simulator is shown in Fig. \ref{Simulator}.

The simulator reads the instructions of the RV32I program binary in Executable and Linkable Format (ELF) from the memory and outputs the instruction numbers and instruction traces as the instructions finish the simulation. The simulator also outputs the performance statistics. 

The simulator does not support wrong-path execution. Although it simulates flushing of the pipeline on branch misprediction, the instruction flushed is the correct instruction.

\subsection{Functional Simulator}
The functional simulator reads the instructions of the RV32I program binary in ELF from the memory. There are 40 unprivileged instructions and six privileged instructions in RV32I. Out of these, the functional simulator updates the state of the program (x registers, program counter and control and status registers) and the state of the hardware (virtualization mode and privilege encoding) on encountering 38 unprivileged instructions (leaving FENCE and EBREAK) and all the six privileged instructions, viz., CSRRW, CSRRS, CSRRC, CSRRWI, CSRRSI and CSRRCI.

\begin{figure}
\centering

\includegraphics[width=0.97\textwidth]{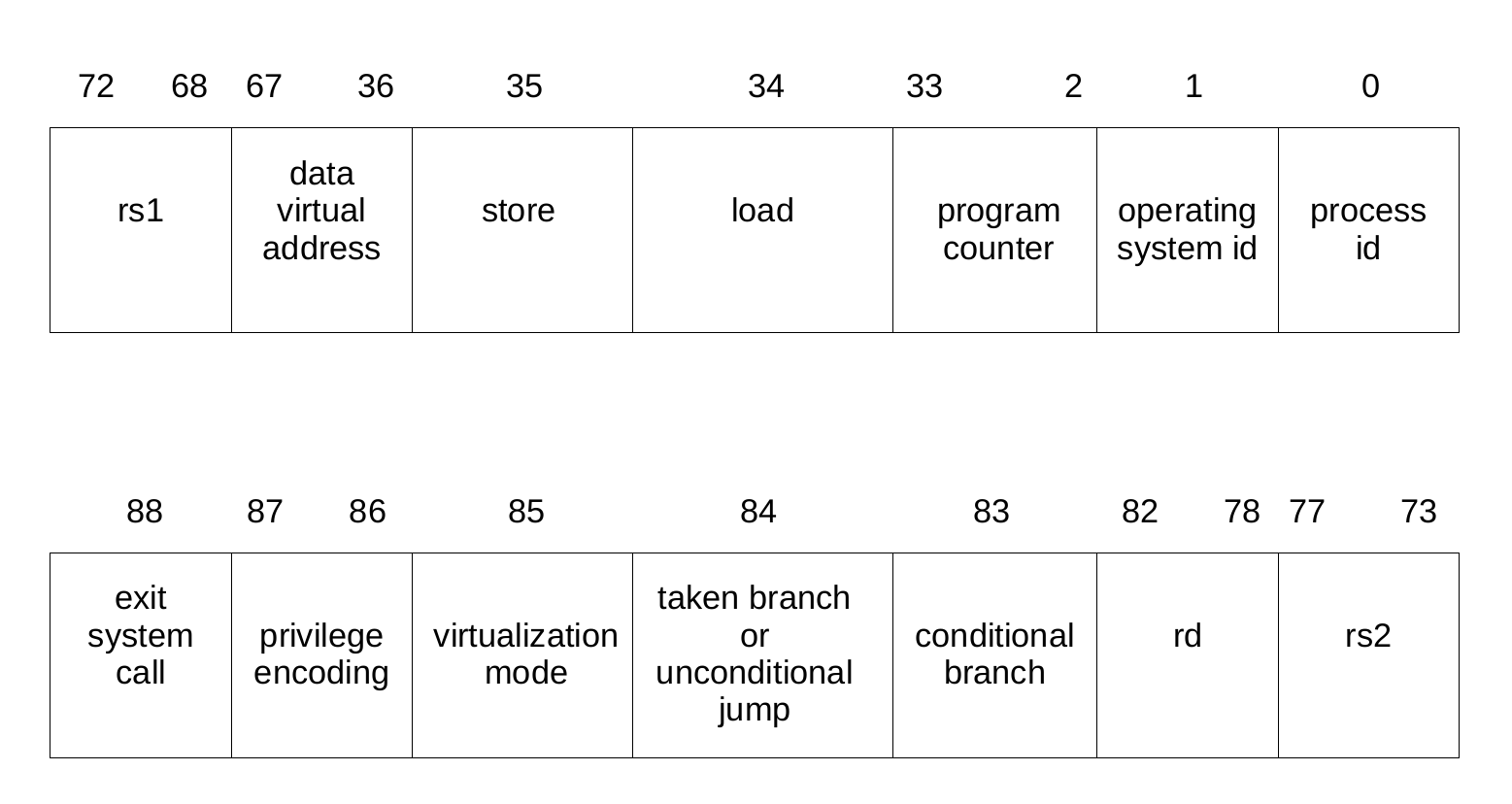}

\caption{Instruction Trace.}
\label{Instruction_Trace}
\end{figure}

The functional simulator uses an instruction counter to assign a distinct instruction number to every dynamic instruction in the RV32I program binary and generates the instruction trace. A \textit{dynamic instruction} is an instruction that the processor sees, as against a \textit{static instruction} which is an instruction that the programmer writes or the compiler or assembler generates. For example, if an instruction is in a loop that is executed multiple times, although it is a single static instruction in the program code, the processor executes it multiple times as multiple dynamic instructions. The format of the instruction trace generated by the functional simulator is shown in Fig. \ref{Instruction_Trace}.

The instruction trace contains process id, operating system id, program counter, data virtual address, rs1 (source register 1), rs2 (source register 2), rd (destination register), virtualization mode and privilege encoding. It also indicates whether the instruction is a load instruction, a store instruction, a conditional branch, a taken branch or unconditional jump or an exit system call by setting the corresponding bits. Currently, the simulator considers only one user process and only one OS. Hence, process id and operating system id are kept as 0. As more user processes and OSs are considered, they may be numbered accordingly using process id and operating system id respectively. Virtualization mode and privilege encoding together indicate the state of the hardware. Exit system call field in the instruction trace, if 1, signals the timing simulator to stop simulating.

\subsection{Timing Simulator}
The timing simulator reads the instruction numbers and instruction traces generated by the functional simulator and outputs the performance statistics along with the instruction numbers and instruction traces that finish the timing simulation. The performance statistics consists of the total number of clock cycles required for the execution of the program and the number of cache and translation lookaside buffer (TLB) misses encountered during the execution of the program. To obtain the performance statistics, the timing simulator simulates \textit{caches}, \textit{TLBs} and \textit{pipeline} and uses a clock cycle counter to count the number of clock cycles.

The timing simulator simulates split cache structure, i.e., there is a separate instruction cache (I-cache) and a separate data cache (D-cache). The cache organization is direct-mapped and the data cache is write-through. The caches are physically-indexed, i.e., the address translation is done before accessing the caches.

Like caches, there is a separate instruction TLB (I-TLB) and data TLB (D-TLB) and the TLB organization is direct-mapped. However, unlike caches, as usual, the TLBs are virtually-indexed, i.e., the address translation is done after accessing the TLBs. The TLB is hardware-managed, i.e., the TLB is managed by the hardware and not by the execution of any software instructions in the ISA. The blocks of main memory, caches and TLBs are assumed to consist of a single word of 32 bits.

The timing simulator simulates a scalar sequential pipeline with pipeline locking. Scalar pipeline \cite{das2014} means that there is a single pipeline and hence, only one instruction is issued to the pipeline at a time. Sequential pipeline \cite{das2014} means that the instructions are completed in-order, i.e., in the order in which they occur in the program. Pipeline locking \cite{Rajaraman2003} means that if an instruction does not advance to the next stage in a particular cycle, all the instructions following the instruction in the pipeline are also prevented from advancing to the next stage in that clock cycle. The simulator simulates a 5-stage pipeline consisting of the stages - instruction fetch (IF), instruction decode (ID), execute (EX), memory access (MEM) and write-back (WB). The number of clock cycles required by the IF stage are determined depending upon whether there is I-TLB miss, I-cache miss and D-cache miss. Similarly, the number of clock cycles required by the MEM stage are determined depending upon whether there is D-TLB miss and D-cache miss. The number of clock cycles required by the ID, EX and WB stages are fixed. The timing simulator simulates memory virtualization by simulating the shadow page table to store the mapping from the virtual address to the host physical address. The timing simulator contains a \textit{hazard detection unit} that simulates structural hazards, load-use data hazards and control hazards. It simulates \textit{forwarding} or \textit{bypassing} \cite{das2014, Patterson2013} that retrieves the missing data element from the internal buffers and reduces the number of data hazards. The simulator simulates branch execution moved from the MEM stage to the ID stage and handles the data hazard arising out of it. The simulator simulates a simple form of branch prediction - assume branch not taken.

The simulator is configured with a set of parameters and the results are obtained by running RV32I program binaries by selecting virtualized and non-virtualized systems.

\section{Simulator configuration}
The simulator is configured with 16 KiB I-cache and 16 KiB D-cache such that each of the cache miss penalty is 100 clock cycles. The main memory write also takes 100 clock cycles. Blocks consist of a single word and each word is of 4 bytes (32 bits). Hence, the number of blocks in each of the I-cache and the D-cache is 4096. Each of the I-TLB and the D-TLB contains 16 page table entries (PTEs) and each PTE is of 4 bytes. Each page is of 4 KiB and hence, the page offset is 12-bit. The virtual address and the host physical address are each 32-bit. If the simulator is configured for virtualized system, in addition to the virtual address and the host physical address, the guest physical address is also 32-bit. The number of clock cycles required by the ID, EX and WB stages are fixed to be one each. The parameters of the simulator along with their values are tabulated in Table \ref{Parameters}.

\begin{table}
\tiny
   \caption{Parameters}
   \label{Parameters}
   \begin{tabular}{ll}
      \hline 
      \multicolumn{1}{c}{Parameter} & \multicolumn{1}{c}{Value}\\\\
      \hline 
      I-cache miss penalty                           & 100 clock cycles\\
      D-cache PTE read miss penalty                  & 100 clock cycles\\
      D-cache data read miss penalty                 & 100 clock cycles\\
      D-cache write miss penalty                     & 100 clock cycles\\
      Clock cycle count for writing into main memory & 100\\\\

      Number of blocks in the I-cache                & 4096\\
      Number of blocks in the D-cache                & 4096\\
      Number of blocks in the I-TLB                  & 16\\
      Number of blocks in the D-TLB                  & 16\\\\

      Data size                                      & 32 bits\\
      Byte offset size                               & 2 bits\\\\

      Page offset size                               & 12 bits\\
      PTE size                                       & 4 bytes\\\\
      
      Virtual address size                           & 32 bits\\
      M-mode virtual address range                   & 0xC0000000 to 0xFFFFFFFF\\
      HS-mode virtual address range                  & 0x80000000 to 0xBFFFFFFF\\
      VS-mode virtual address range                  & 0x40000000 to 0x7FFFFFFF\\
      U-mode / VU-mode virtual address range         & 0x00000000 to 0x3FFFFFFF\\\\

      Guest physical address size                    & 32 bits\\
      M-mode guest physical address range            & 0x00000000 to 0x3FFFFFFF\\
      HS-mode guest physical address range           & 0x40000000 to 0x7FFFFFFF\\
      VS-mode guest physical address range           & 0x80000000 to 0xBFFFFFFF\\
      U-mode / VU-mode guest physical address range  & 0xC0000000 to 0xFFFFFFFF\\\\

      Host physical address size                          & 32 bits\\
      M-mode host physical address range             & 0x00000000 to 0x3FFFFFFF\\
      HS-mode host physical address range            & 0x40000000 to 0x7FFFFFFF\\
      VS-mode host physical address range            & 0x80000000 to 0xBFFFFFFF\\
      U-mode / VU-mode host physical address range   & 0xC0000000 to 0xFFFFFFFF\\\\
      
      Clock cyle count for ID stage                  & 1\\
      Clock cyle count for EX stage                  & 1\\
      Clock cyle count for WB stage                  & 1\\\\
      
      Parameter to select between                    & 0: Non-virtualized system\\
      virtualized and non-virtualized systems        & 1: Virtualized system\\\\
      \hline 
   \end{tabular}
\end{table}

Two user programs - one for searching (linear search) and one for sorting (bubble sort) are fed as input to the simulator. Both the programs use write system calls to output the result to the standard output (console). For each of the two programs, the simulator is configured first for non-virtualized system and then for virtualized system and the results are obtained.

\section{Results}
The performance statistics obtained from the simulator while running searching program (without virtualization and with virtualization) are tabulated in Table \ref{Performance_statistics_obtained_from_the_simulator_while_running_searching_program_(without_and_with_virtualization)}. Likewise, the performance statistics obtained while running sorting program are tabulated in Table \ref{Performance_statistics_obtained_from_the_simulator_while_running_sorting_program_(without_and_with_virtualization)}. The percentage overhead in the performance statistics due to virtualization is also given for both the programs.

\begin{table}
\tiny
   \caption{Performance statistics obtained from the simulator while running searching program (without and with virtualization)}
   \label{Performance_statistics_obtained_from_the_simulator_while_running_searching_program_(without_and_with_virtualization)}
   \begin{tabular}{lccc}
      \hline 
                                                 & Without        & With           & Percentage\\
                                                 & Virtualization & Virtualization & Overhead\\
                                                 &                &                & Due to\\
      \multicolumn{1}{c}{Performance Statistics} &                &                & Virtualization\\\\
                                                 & (N)            & (V)            & $ \frac{V-N}{N} \times 100 \% $\\\\
      \hline 
      Number of I-TLB misses while reading PTE in IF stage                  & 6       & 10      & 40 \%\\
      Number of I-cache misses while reading instruction in IF stage        & 402     & 426     & 5.634 \%\\
      Number of D-cache misses while reading PTE in IF stage                & 3       & 7       & 57.14 \%\\ 
      Number of D-TLB misses while reading PTE in MEM stage during load     & 9       & 11      & 18.18 \%\\
      Number of D-cache misses while reading data in MEM stage during load  & 39      & 39      & 0\\   
      Number of D-cache misses while reading PTE in MEM stage during load   & 8       & 10      & 20 \%\\
      Number of D-TLB misses while reading PTE in MEM stage during store    & 4       & 6       & 33.33 \%\\
      Number of D-cache misses while writing data in MEM stage during store & 65      & 67      & 40.00 \%\\
      Number of D-cache misses while reading PTE in MEM stage during store  & 4       & 6       & 2.985 \%\\\\
      
      Total number of I-TLB misses                                          & 6       & 10      & 40 \%\\
      Total number of I-cache misses                                        & 402     & 426     & 5.634 \%\\
      Total number of D-TLB misses                                          & 13      & 17      & 23.53 \%\\
      Total number of D-cache misses                                        & 119     & 129     & 7.752 \%\\
      Clock cycle count                                                     & 65153   & 68322   & 4.638 \%\\
      Number of dynamic instructions executed (simulated)                   & 894     & 948     & 5.696 \%\\
      Cycles per instruction (CPI)                                          & 72.88   & 72.07   & -1.124 \%\\
      Instructions per cycle (IPC)                                          & 0.01372 & 0.01388 & 1.153 \%\\
      \hline 
   \end{tabular}
\end{table}

\begin{table}
\tiny
   \caption{Performance statistics obtained from the simulator while running sorting program (without and with virtualization)}
   \label{Performance_statistics_obtained_from_the_simulator_while_running_sorting_program_(without_and_with_virtualization)}
   \begin{tabular}{lccc}
      \hline 
                                                 & Without        & With           & Percentage\\
                                                 & Virtualization & Virtualization & Overhead\\
                                                 &                &                & Due to\\
      \multicolumn{1}{c}{Performance Statistics} &                &                & Virtualization\\\\
                                                 & (N)            & (V)            & $ \frac{V-N}{N} \times 100 \% $\\\\
      \hline 
      Number of I-TLB misses while reading PTE in IF stage                  & 40      & 78      & 48.72 \%\\
      Number of I-cache misses while reading instruction in IF stage        & 424     & 448     & 5.357 \%\\
      Number of D-cache misses while reading PTE in IF stage                & 3       & 41      & 92.68 \%\\ 
      Number of D-TLB misses while reading PTE in MEM stage during load     & 60      & 79      & 24.05 \%\\
      Number of D-cache misses while reading data in MEM stage during load  & 25      & 25      & 0\\   
      Number of D-cache misses while reading PTE in MEM stage during load   & 59      & 78      & 24.36 \%\\
      Number of D-TLB misses while reading PTE in MEM stage during store    & 21      & 40      & 47.50 \%\\
      Number of D-cache misses while writing data in MEM stage during store & 55      & 57      & 3.509 \%\\
      Number of D-cache misses while reading PTE in MEM stage during store  & 21      & 40      & 47.50 \%\\\\
      
      Total number of I-TLB misses                                          & 40      & 78      & 48.72 \%\\
      Total number of I-cache misses                                        & 424     & 448     & 5.357 \%\\
      Total number of D-TLB misses                                          & 81      & 119     & 31.93 \%\\
      Total number of D-cache misses                                        & 163     & 241     & 32.37 \%\\
      Clock cycle count                                                     & 98163   & 108744  & 9.730 \%\\
      Number of dynamic instructions executed (simulated)                   & 3324    & 3837    & 13.37 \%\\
      Cycles per instruction (CPI)                                          & 29.53   & 28.34   & -4.199 \%\\
      Instructions per cycle (IPC)                                          & 0.03386 & 0.03528 & 4.025 \%\\
      \hline 
   \end{tabular}
\end{table}

Cycles per instruction (CPI) is calculated by dividing the clock cycle count by the number of dynamic instructions executed (simulated). CPI is much higher than 1 because the block size is taken to be one word, thereby, exploiting only the temporal locality and not the spatial locality of caches. Furthermore, only a single-level cache is simulated and the data cache is write-through. It is seen that the number of instructions executed in a virtualized system are higher than the number of instructions executed in a non-virtualized system. This is due to the hypervisor instructions that get executed in a virtualized system. It is seen that the CPI with virtualization is less than the CPI without virtualization and the overhead in CPI due to virtualization is negative. This is because of the nature of pipelining. As the number of instructions executed increases, the CPI in a pipelined architecture reduces.

\section{Validation}
The validation of the functional simulator is done by rigourous checking of the state of the program (registers and memory) after the simulation of the instructions. Particularly, it is verified that the data in the U-mode/VU-mode address space is correctly transferred to the standard output port address when a write system call is executed.

The validation of the timing simulator is done by tracing whether the instructions correctly enter into the pipeline and verifying that the pipeline hazards are correctly simulated. The performance statistics are calculated for the initial few instructions and they are compared with those output by the simulator and are found to be correct. As expected, the number of instructions executed in a virtualized system are more than those executed in a non-virtualized system and the CPI reduces as more number of instructions are executed in a pipelined architecture.

\section{Conclusion}
Virtualization has regained importance and this has put a pressing demand for the simulation of virtualized systems. In order to accurately simulate the overhead due to virtualization, it is necessary for a full-system simulator to simulate the hypervisor instructions in addition to the operating system instructions. However, this increases the simulation time. Hence, this paper proposes a hardware-based approach for the simulation of hypervisor instructions because hardware-based simulator is faster than the software-based simulator and can simulate the increased overhead due to virtualization in reasonable time.

The proposed approach is demonstrated by simulating RISC-V program binary. Particularly, the base integer ISA, RV32I, of RISC-V is simulated. The simulator is coded in Verilog so as to develop a hardware-based simulator. The simulator is a decoupled simulator, particularly, functional-first simulator and is an execution-driven simulator. The functional simulator maintains the state of the program and the state of the hardware and generates the instruction trace from the RV32I program binary. The timing simulator accepts the instruction trace generated by the functional simulator, inserts it into its simulated pipeline and generates the performance statistics. The performance statistics consists of the total number of clock cycles required for the execution of the program and the number of cache and TLB misses encountered during the execution of the program.

The proposed approach simulates hypervisor instructions which leads to accurate timing simulation of virtualized systems.

\section{Future scope}
The proposed approach may be extended to simulate multicore processors. The organization of the hardware components of the simulator so as to maximize parallelism is also an interesting research direction to reduce the simulation time.\\

\textbf{Funding:}
This research did not receive any specific grant from funding agencies in the public, commercial, or not-for-profit sectors.

\small
\bibliography{1-base}

\end{document}